\documentclass[a4paper,fleqn]{cas-dc}

\usepackage[english]{babel}
\usepackage[numbers,sort&compress]{natbib}
\usepackage{graphicx}
\usepackage{graphics}
\usepackage{braket}
\usepackage{bbold}
\usepackage{amssymb}
\usepackage{amsmath}
\usepackage{nicefrac}
\usepackage{dcolumn}% Align table columns on decimal point
\usepackage{bm}% bold math
\usepackage{slashed}
\usepackage{datetime}
\usepackage{mciteplus}
\usepackage{multirow}
\usepackage{bbold}
\usepackage{color}
\usepackage{xcolor}
\usepackage{float}
\usepackage[utf8]{inputenc}
\usepackage[normalem]{ulem}

\def\tsc#1{\csdef{#1}{\textsc{\lowercase{#1}}\xspace}}
\tsc{WGM}
\tsc{QE}
\newcommand{\DY}{\Delta Y}
\newcommand{\JPsi}{J/\psi}

\newcommand{\tref}[1]{~\ref{#1}}

\begin{document}
\let\WriteBookmarks\relax
\def\floatpagepagefraction{1}
\def\textpagefraction{.001}

\shorttitle{Stabilizing BFKL via heavy-flavor and NRQCD fragmentation}    

\shortauthors{Celiberto, Francesco Giovanni}  

\title []{\Huge Stabilizing BFKL via heavy-flavor \\ and NRQCD fragmentation}  

\author[1,2,3]{Francesco Giovanni Celiberto}[orcid=0000-0003-3299-2203]

\cormark[1]

%\fnmark[<footnote mark no>]

\ead{fceliberto@ectstar.eu}

%\ead[url]{<URL>}

\affiliation[1]{organization={European Centre for Theoretical Studies in Nuclear Physics and Related Areas (ECT*)},
            addressline={Strada delle Tabarelle 286}, 
            city={Villazzano},
%          citysep={}, % Uncomment if no comma needed between city and postcode
            postcode={I-38123}, 
            state={Trento},
            country={Italy}}

\affiliation[2]{organization={Fondazione Bruno Kessler (FBK)},
            addressline={Via Sommarive 18}, 
            city={Povo},
%          citysep={}, % Uncomment if no comma needed between city and postcode
            postcode={I-38123}, 
            state={Trento},
            country={Italy}}

\affiliation[3]{organization={INFN-TIFPA Trento Institute of Fundamental Physics and Applications},
            addressline={Via Sommarive 14}, 
            city={Povo},
%          citysep={}, % Uncomment if no comma needed between city and postcode
            postcode={I-38123}, 
            state={Trento},
            country={Italy}}

%\cortext[1]{Corresponding author}



%-----------------------------------------
\begin{abstract}
We bring evidence that the recently discovered property of natural stability of the high-energy resummation is directly connected to the fragmentation mechanism of heavy hadrons. As a phenomenological support, we provide predictions for differential distributions sensitive to heavy-hadron tags, calculated at the next-to-leading logarithmic level of the hybrid high-energy/collinear factorization (NLL/NLO), as implemented in the {\tt JETHAD} multimodular code. We show that the stabilizing mechanism is encoded in gluon channels of both heavy-flavor collinear fragmentation functions extracted from data and the ones evolved from a non-relativistic QCD input.
\end{abstract}
%-----------------------------------------


%\item 
%\item 
%\item 

% Keywords
% Each keyword is separated by \sep

\maketitle

\section{Hors d'{\oe}uvre}
\label{sec:intro}

It is widely known that high-energy calculations via the next-to-leading BFKL resummation~\cite{Fadin:1975cb,Balitsky:1978ic} of energy logarithms (NLL) suffer from strong instabilities that become manifest when renormalization and factorization scales are varied from their natural values, namely the ones dictated by kinematics. For observables featuring light-particle emissions, such as Mueller--Navelet~\cite{Mueller:1986ey,Colferai:2010wu,Ducloue:2013hia,Ducloue:2013bva,Caporale:2014gpa,Celiberto:2015yba,Celiberto:2015mpa,Celiberto:2016ygs,Celiberto:2016vva,Caporale:2018qnm,Celiberto:2022gji} and light-hadron correlations~\cite{Celiberto:2016hae,Celiberto:2016zgb,Celiberto:2017ptm,Celiberto:2017uae,Celiberto:2017ydk,Bolognino:2018oth,Bolognino:2019cac,Bolognino:2019yqj,Celiberto:2020wpk,Celiberto:2020rxb}, these instabilities are so strong to hamper any possibility to perform precision studies at natural scales (for further advancements in high-energy QCD phenomenology, see~\cite{Boussarie:2017oae,Golec-Biernat:2018kem,Celiberto:2017nyx,Bolognino:2019ouc,Bolognino:2019yls,Bolognino:2019ccd,Celiberto:2021dzy,Celiberto:2021fdp,Bolognino:2022wgl,Celiberto:2022dyf,Celiberto:2022keu,Celiberto:2022zdg,Celiberto:2022kza,Bolognino:2021mrc,Bolognino:2021hxxaux,Caporale:2015vya,Caporale:2015int,Caporale:2016soq,Caporale:2016vxt,Caporale:2016xku,Celiberto:2016vhn,Caporale:2016djm,Caporale:2016pqe,Chachamis:2016qct,Chachamis:2016lyi,Caporale:2016lnh,Caporale:2016zkc,Chachamis:2017vfa,Caporale:2017jqj,Celiberto:2020tmb,Celiberto:2021fjf,Celiberto:2021tky,Celiberto:2021txb,Celiberto:2021xpm,Hentschinski:2020tbi,Celiberto:2022fgx,Celiberto:2022qbh}). We provide evidence that LHC final states sensitive to heavy-flavored hadrons exhibit a fair and solid stability of these observables under higher-order corrections and scale variations. The stabilization mechanism is encoded in the peculiar behavior of the gluon collinear fragmentation function (FF) describing the heavy hadron. It comes out as is an intrinsic feature emerging whenever a species with heavy flavor is detected. This remarkable property, called \emph{natural stability} of the high-energy resummation, is typical of both heavy-flavor FFs extracted from data and the ones evolved from a non-relativistic QCD (NRQCD) input~\cite{Caswell:1985ui,Thacker:1990bm,Bodwin:1994jh}.

\section{Natural stability from fragmentation}
\label{sec:stability}

\begin{figure*}[t]
\centering

   \includegraphics[scale=0.55,clip]{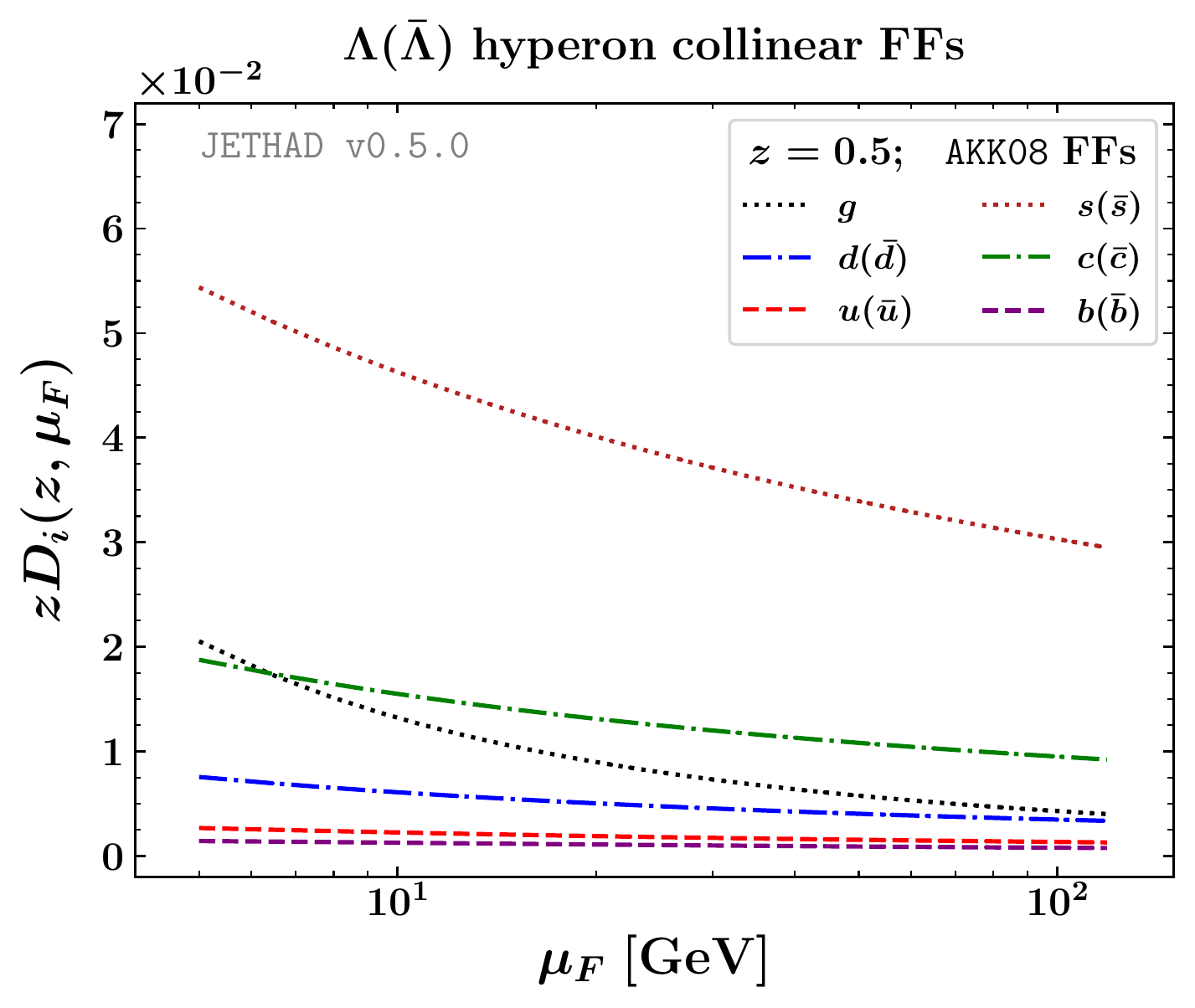}
   \hspace{0.50cm}
   \includegraphics[scale=0.46,clip]{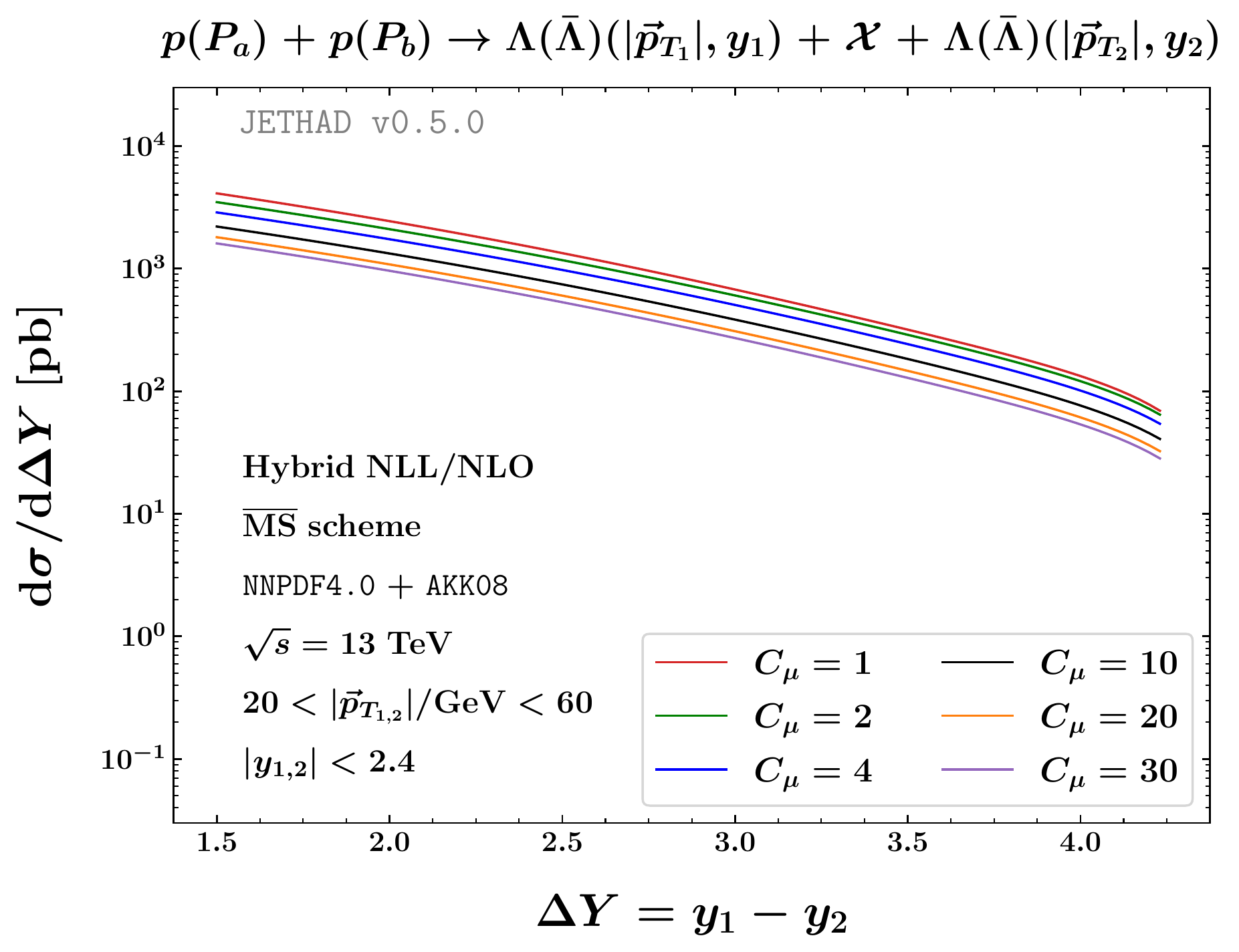}

   \includegraphics[scale=0.55,clip]{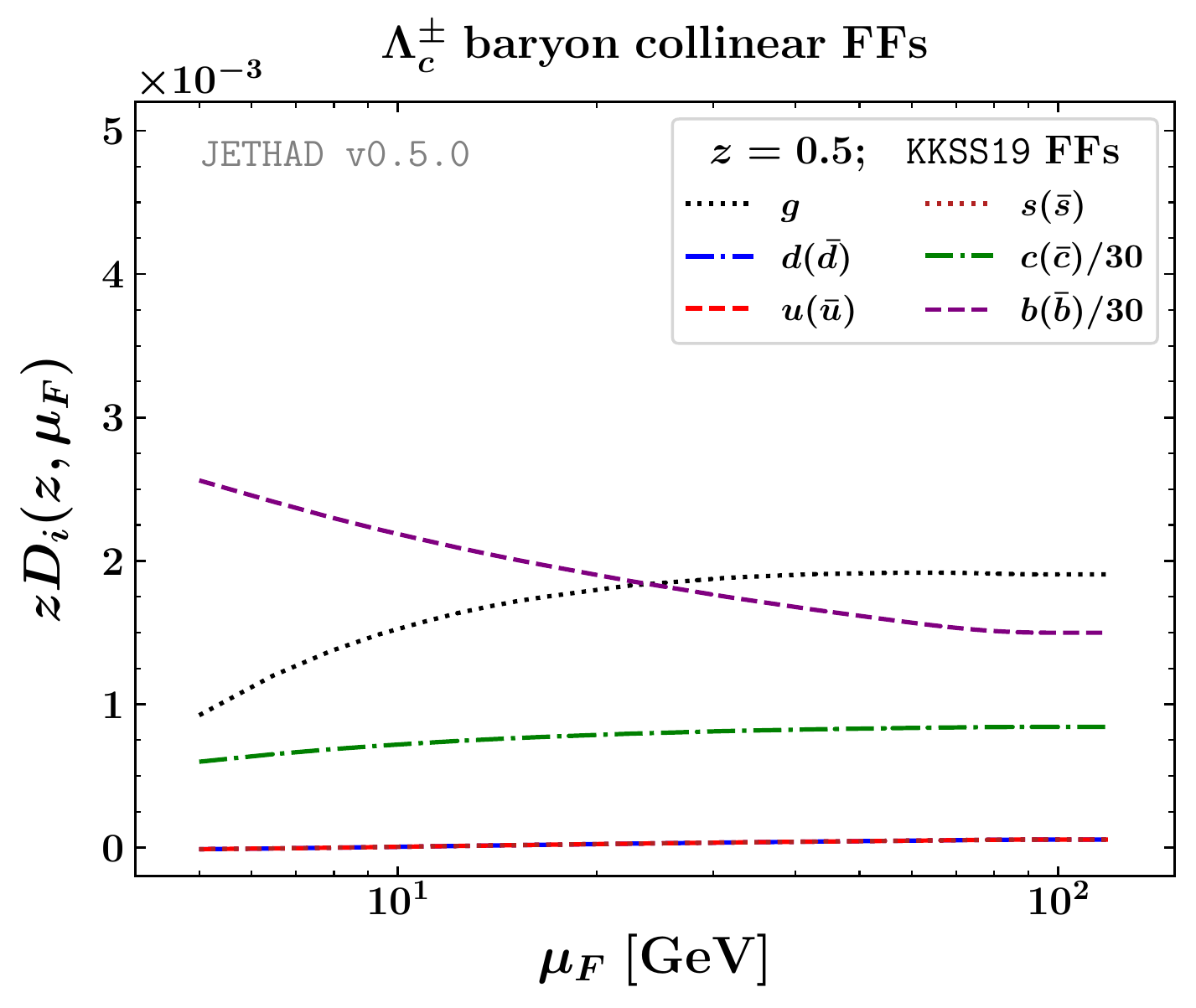}
   \hspace{0.50cm}
   \includegraphics[scale=0.46,clip]{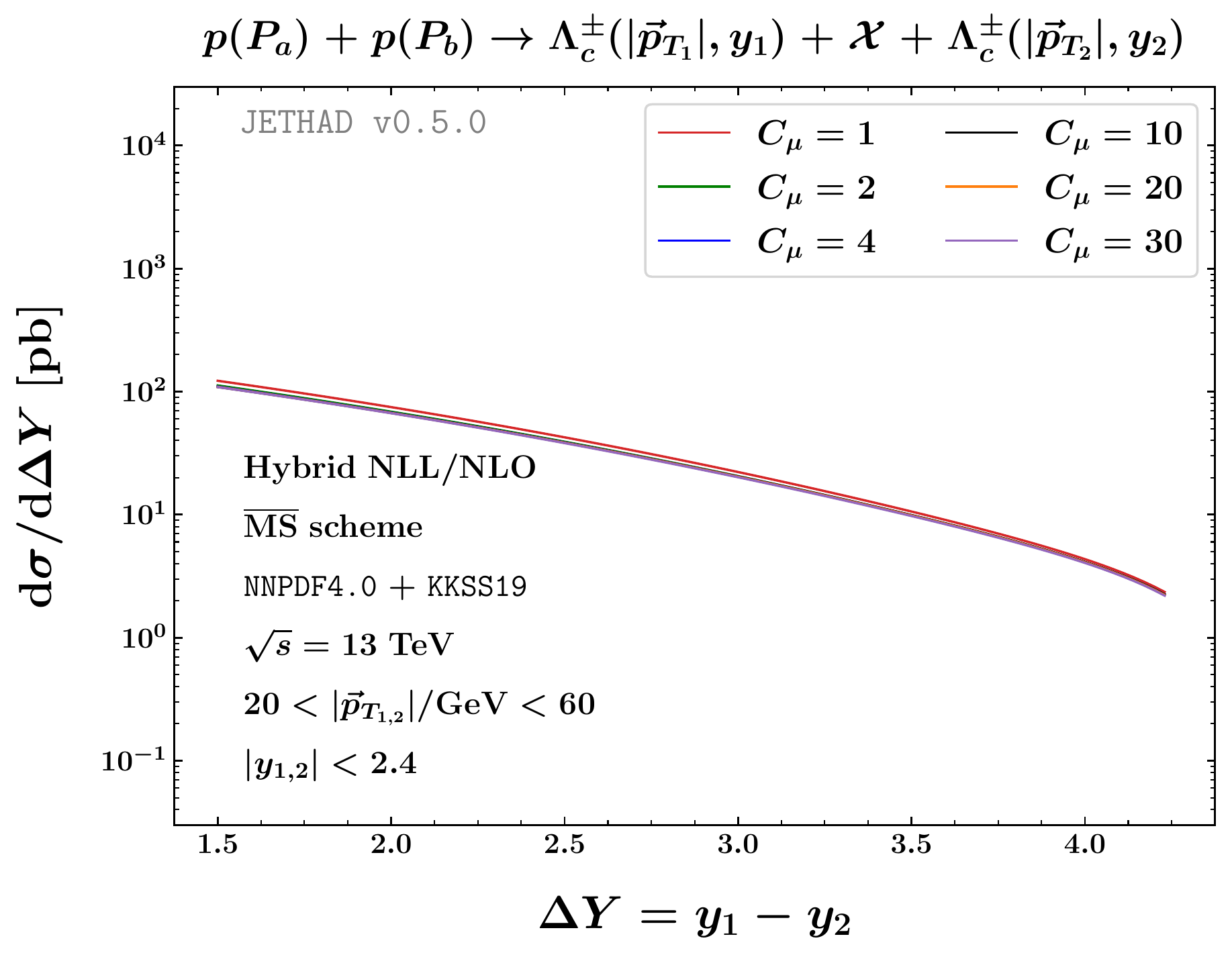}

   \includegraphics[scale=0.55,clip]{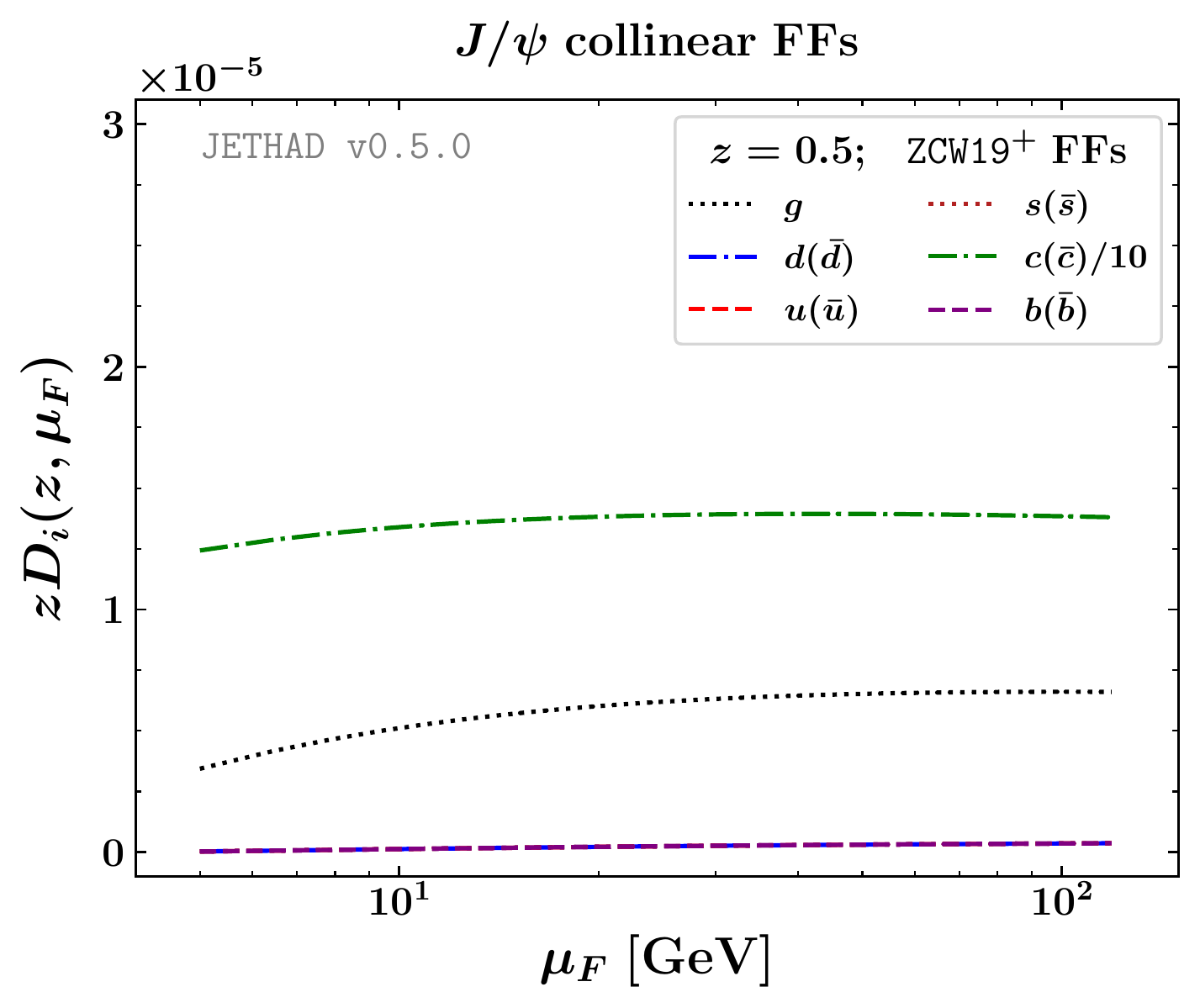}
   \hspace{0.50cm}
   \includegraphics[scale=0.46,clip]{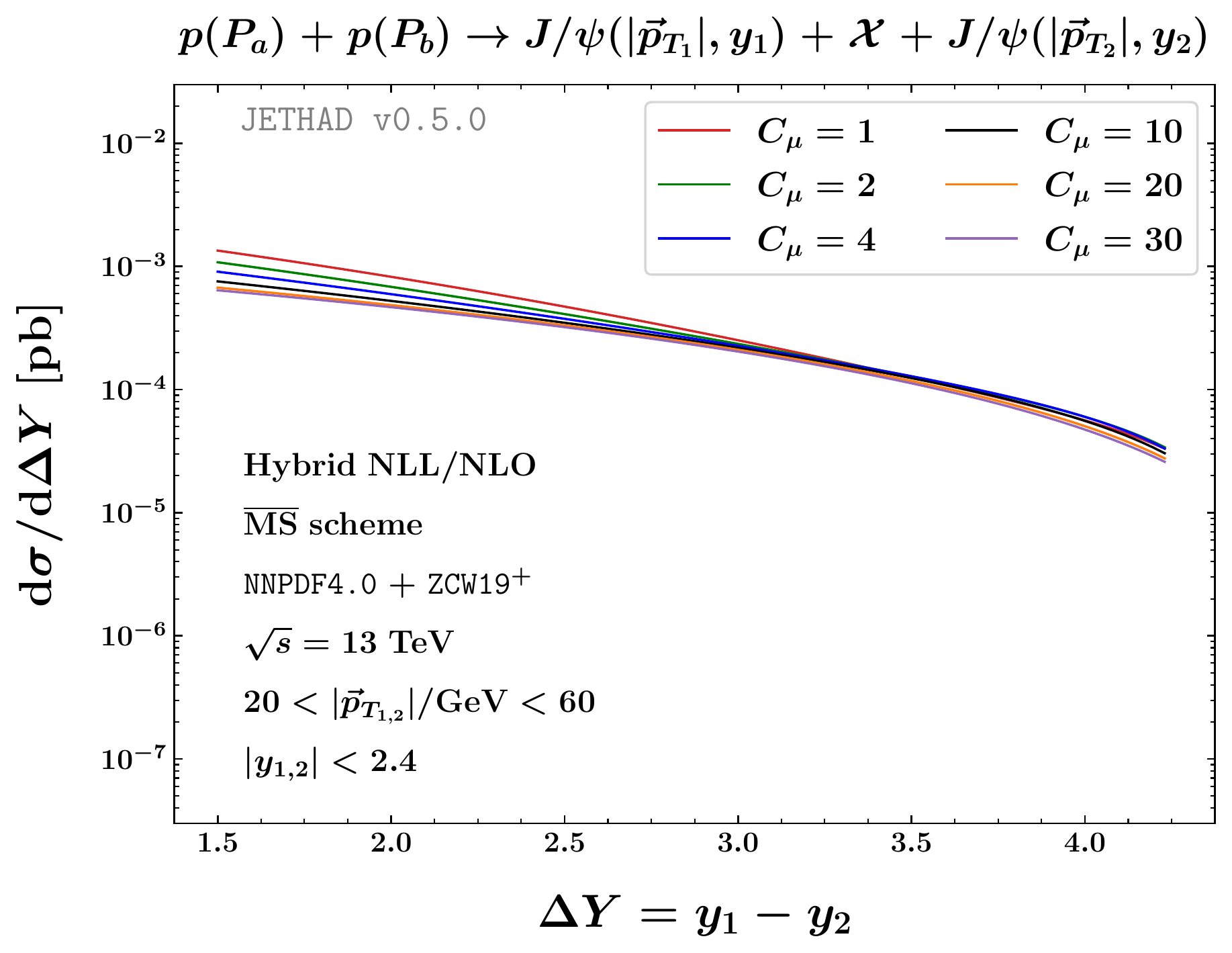}

\caption{Left panels: $\mu_F$-dependence of: $\Lambda (\bar \Lambda)$ {\tt AKK08} (upper), $\Lambda_c^{\pm}$ {\tt KKSS19} (central), and $\JPsi$ {\tt ZCW19$^+$} (lower) NLO FFs at $z = 5 \times 10^{-1}$.
Right panels: rapidity distribution in dihadron channels at $\sqrt{s} = 13$ TeV.}
\label{fig:FFs_Y_psv}
\end{figure*}

We study inclusive dihadron tags at the LHC within the NLL/NLO hybrid factorization as implemented in {\tt JETHAD}~\cite{Celiberto:2020wpk,Celiberto:2022rfj}. Left panels of Fig.\tref{fig:FFs_Y_psv} show the $\mu_F$-dependence of: $\Lambda (\bar \Lambda)$ {\tt AKK08}~\cite{Albino:2008fy}, $\Lambda_c^{\pm}$ {\tt KKSS19}~\cite{Kniehl:2020szu}, and $\JPsi$ {\tt ZCW19$^+$}~\cite{Celiberto:2022dyf} NLO FFs at $z = 5 \times 10^{-1}$. This roughly corresponds to the average value the momentum fraction, $\langle z \rangle$, at which FFs are typically probed by rapidity distributions at LHC. They are cross sections differential in the rapidity interval, $\DY$, between the two hadrons, see right panels of Fig.\tref{fig:FFs_Y_psv}. NLO parton distribution functions (PDFs) are taken from {\tt NNPDF4.0}~\cite{NNPDF:2021njg}.
We clearly observe that the $\DY$ distribution is very sensitive to scale variation ($\mu_R$ and $\mu_R$ range from one to 30 times their natural values, \emph{i.e.} the transverse masses of the final-state objects) in the lighter $\Lambda (\bar \Lambda)$ case, whereas it is very stable when heavier hadrons are tagged.
As pointed out in~\cite{Celiberto:2021dzy,Celiberto:2022dyf}, a key role is played by the heavy-hadron gluon FF. Its impact on the cross section gets enhanced by the convolution with the gluon PDF at leading order, which dominates also over non-diagonal $gq$ channels at NLO.
Smooth-behaved, non-decreasing with $\mu_F$ gluon FFs (central and lower left panels of Fig.\tref{fig:FFs_Y_psv}) compensate the falloff with $\mu_R$ of the running coupling in the
high-energy cross section. This generates the stability observed in heavy-flavor distributions.
This \emph{natural stability} is a general feature emerging whenever a heavy-flavored hadron is emitted, independently of \emph{Ans\"atze} made on corresponding FFs. It holds both for single-charmed hyperons, whose {\tt KKSS19} FF determination was extracted from data, and for vector quarkonia, whose {\tt ZCW19$^+$} FF set was obtained by evolution from an initial-scale calculation~\cite{Braaten:1993mp,Braaten:1993rw,Zheng:2019dfk} within the NRQCD framework.
Analogous results were found for bottomed hadrons~\cite{Celiberto:2021fdp,Celiberto:2022keu}, while milder stabilizing effects were observed for lighter $\Xi^-/\bar\Xi^+$ baryons~\cite{Celiberto:2022kxx}.

\section{Toward precision studies}
\label{sec:conclusions}

By investigating LHC final states featuring emissions of heavy-flavored bound states at high rapidities and transverse momenta, we have provided an indisputable evidence that a \emph{natural-stabilization} pattern emerges and it permits to perform NLL/NLO studies of related differential distribution.
We have discovered that natural stability is a remarkable property shared by all the heavy-flavored species analyzed so far: heavy-light hadrons~\cite{Celiberto:2021dzy,Celiberto:2021fdp,Celiberto:2022rfj,Celiberto:2022zdg}, vector quarkonia~\cite{Celiberto:2022dyf}, and $B_c^{(*)}$ mesons~\cite{Celiberto:2022keu}.
The possibility of studying high-energy QCD at the natural scales provided by process kinematics was a required point and the first milestone to move our first steps toward the precision level.
First, a formal proof of the natural stability, emerged so far as a phenomenological property, needs to be afforded.
Then, the hybrid factorization needs to evolve into \emph{multi-lateral} and unified formalism that where several different resummations are simultaneously embodied.
These are required steps to deepen our knowledge of high-energy QCD at new-generation colliding machines~\cite{AbdulKhalek:2021gbh,Khalek:2022bzd,Chapon:2020heu,Begel:2022kwp,Dawson:2022zbb,Bose:2022obr,Anchordoqui:2021ghd,Feng:2022inv,Hentschinski:2022xnd,Acosta:2022ejc,AlexanderAryshev:2022pkx,Brunner:2022usy,Arbuzov:2020cqg,Amoroso:2022eow,Black:2022cth,MuonCollider:2022xlm,Aime:2022flm,MuonCollider:2022ded} and access the proton structure at low $x$ via the gluon distributions~\cite{Besse:2013muy,Bolognino:2018rhb,Bolognino:2018mlw,Bolognino:2019bko,Bolognino:2019pba,Celiberto:2019slj,Bolognino:2021niq,Bolognino:2021gjm,Bolognino:2022uty,Celiberto:2022fam,Bolognino:2022ndh,Brzeminski:2016lwh,Celiberto:2018muu,Garcia:2019tne,Bacchetta:2020vty,Celiberto:2021zww,Bacchetta:2021oht,Bacchetta:2021lvw,Bacchetta:2021twk,Bacchetta:2022esb,Bacchetta:2022crh,Bacchetta:2022nyv,Celiberto:2022omz,Forte:2015gve,Ball:2017otu}.

%-----------------------------------------
\section*{Acknowledgments}
%-----------------------------------------

Results presented in this manuscript rely on the {\tt ZCW19$^+$} vector-quarkonium collinear NLO FF set, built in~\cite{Celiberto:2022dyf} on the basis of initial-scale NRQCD inputs for constituent heavy-quark~\cite{Braaten:1993mp,Zheng:2019dfk} and gluon~\cite{Braaten:1993rw} channels.  
The author thanks Alessandro Papa,  Dmitry Yu. Ivanov, Michael Fucilla, and Mohammed M.A. Mohammed for inspiring conversations.
The author is grateful to Jean-Philippe~Lansberg, Hua-Sheng~Shao, and colleagues of the \emph{Quarkonia As Tools} Community for insightful discussions on production mechanisms and phenomenology of heavy-flavored bound states.
This work was supported by the INFN/NINPHA project.
The author thanks the Universit\`a degli Studi di Pavia for the warm hospitality.

%\bibliographystyle{cas-model2-names}
\bibliographystyle{elsarticle-num}

%-----------------------------------------
\bibliography{bibliography}
%-----------------------------------------

\end{document}